\documentclass[aps,fleqn,preprint]{revtex4-1}

\usepackage{amsmath}
\usepackage{graphicx}
\usepackage{epsfig}
\usepackage{subfigure}
\usepackage{amsfonts}
\usepackage{amssymb}
\usepackage{amsthm}
\usepackage{newlfont}
\usepackage{epstopdf}

\usepackage[utf8]{inputenc}

\usepackage{dcolumn}
\usepackage{bm}

\usepackage[section] {placeins}

\begin{document}

\title{ Calculation of the entropy for hard-sphere from integral equation method  }

\author{Purevkhuu Munkhbaatar$^{1}$}
\author{Banzragch Tsednee$^{1}$}
\author{Tsogbayar Tsednee$^{1,2}$}
\author{Tsookhuu Khinayat$^{1}$}
\affiliation{$^1$Institute of Physics and Technology, Mongolian Academy of Sciences, Peace Ave 54B, Ulaanbaatar 13330, Mongolia, $^{2}$Chemistry Department, University of North Dakota, Grand Forks, ND 58202, USA}

\begin{abstract}

The Ornstein-Zernike integral equation method has been employed for a single-component hard sphere fluid in terms of the Percus-Yevick (PY) and Martynov-Sarkisov (MS) approximations.  Virial equation of state has been computed in both approximations. An excess chemical potential has been calculated with an analytical expression based on correlation functions, and the entropy has been computed with a thermodynamic relation. Calculations have been carried out for a reduced densities of 0.1 to 0.9. It has been shown that the MS approximation gives better values than those from the PY approximation, especially for high densities and presents a reasonable comparison with available data in the literature.

\end{abstract}

\pacs{Valid PACS appear here}
                             

\maketitle

\section{Introduction}

In classical statistical physics the physical systems, such as, a liquid can be described  as a spherical symmetric hard-sphere particle model or Lennard-Jones potentials \cite{Hansen06, McQuarried10}. Theoretical investigations for the systems can be performed with various methods, such as an explicit approaches--Monte-Carlo or Molecular dynamics simulations \cite{Hansen06}, and an implicit approaches represented as an integral equation or a polarizable continuum model methods \cite{Hirataed03}. An integral equation (IE) approach mentioned here is a mathematical tool we use in our study. As an implicit approach, the IE method does not consider a number of particles in the system which in turn may make a calculation a cheap, and a solution of the IE can give directly the correlation functions determining a general structure of the system.  A one-component, homogeneous system can be successfully investigated with the Ornstein-Zernike (OZ) \cite{OrnsteinZ14} IE theory combined with an appropriate auxiliary equation. 

In this work our purpose is to obtain an excess entropy for single-component hard-sphere fluid using the OZ  IE approach combined with the Percus-Yevick~\cite{PYpr68} and Martynov-Sarkisov \cite{MSmp68} approximations. To reach it, we will first compute a virial equation of state and along with it, we will obtain an excess chemical potential for the system at equilibrium using an analytical expression based on the correlation function. Using these two quantities, we will compute the excess entropy employing a thermodynamic relation. We will compare our findings for thermodynamic properties with available data in literature \cite{Zhou06, Carnahan69}.  Note that, to our knowledge, the MS approximation has not been tested for this calculation, yet, even though this problem had been considered in the past \cite{Zhou06, Carnahan69, Laird92}.
 Therefore, we believe that our findings in this work can be considered as some contributions to this area.  

In Section II we will discuss about the Ornstein-Zernike theory and thermodynamic properties which we will compute.  In Section III we present numerical results and their discussions. In Section IV the conclusion is given.   

\section{Theory}

\subsection{The Ornstein-Zernike equation} 	

In statistical mechanics an structural properties for the liquids existing at equilibrium can be obtained in terms of the integral equation formalism \cite{Hirataed03}. For a single-component homogeneous system the Ornstein-Zernike integral equation can be written in the form
\begin{eqnarray}\label{oz}
 h(r) = c(r) + \rho \int c(|\mathbf{r} - \mathbf{r}'|) h(\mathbf{r}') d\mathbf{r}'
 \end{eqnarray}
where $h(r)$ and $c(r)$ are the total and direct correlation functions, respectively, and $\rho$ is density of the system. 
 
In equation (\ref{oz}), the two correlation functions are unknown, therefore, it cannot be solved directly. To solve this OZ equation (\ref{oz}), we need another equation which is solved together with the OZ equation (\ref{oz}) in the self-consistent way. This required equation is called a {\it closure equation (relation)} which may be written in the form 
\begin{eqnarray}\label{cl}
 h(r) = \exp[-\beta u(r) + \gamma(r) - B(r)] - 1,
 \end{eqnarray}
where $u(r)$ is a pair potential for particles in the system; $\gamma (r) = h(r) - c(r)$ is an indirect correlation function; $B(r)$ is a bridge function; $\beta = 1/k_{\mathrm{B}}T$ where $k_{\mathrm{B}}$ is the Boltzmann's constant and $T$ is a temperature of the system.  The radial distribution function $g$ can be defined as $g(r) = h(r) + 1$ as well. 

An inter-particle interaction potential for the hard-sphere with a diameter $\sigma$ can be written in the form \cite{McQuarried10}
\begin{eqnarray}\label{hs}
 u(r) =  
 \begin{cases}
  0, & r\leq \sigma\\
  \infty, & r > \sigma. 
 \end{cases}
 \end{eqnarray}
 
 The bridge functions we use in this work can be given in the forms:
\begin{eqnarray}\label{br}\nonumber
 B(r) =  
 \begin{cases}
 \ln(1+\gamma) - \gamma, & \mbox{Percus-Yevick (PY)}\,\, [5] \\
   \sqrt{1+ 2\gamma} -\gamma -1, & \mbox{Martynov-Sarkisov (MS)}\,\, [6]. 
 \end{cases}
\end{eqnarray}  
  
\subsection{Thermodynamic quantities}

Once we solve the OZ equation (\ref{oz}), we can have the correlation functions with which we can compute thermodynamic properties for the system. 

\subsubsection{Virial equation of state}

For the hard-sphere system, the virial equation of state can be written in the form \cite{McQuarried10}
\begin{eqnarray}\label{veos}
 \frac{\beta p}{\rho} = 1 + \frac{2 \pi}{3} \rho \sigma^{3} g(\sigma), 
\end{eqnarray}
where $g(\sigma)$ is the contact value of the radial distribution function at $\sigma$.  

\subsubsection{An excess chemical potential}

The excess (e) chemical potential $\beta \mu^{e}$ can be computed with a following approximated analytical expression
\begin{eqnarray}\label{veos}
  \beta \mu^{e} & \approx &  \rho \int \Big(\frac{1}{2}h(r)^{2} - c(r) - \frac{1}{2} h(r) c(r) \Big) d\mathbf{r}  \\ \nonumber
  & & + \rho \int \Big( B (r) + \frac{2}{3} h(r) B(r)\Big) d\mathbf{r}.
\end{eqnarray}
Note that this expression does not depend an explicit form of the bridge function. Therefore, we can use it for both approximation in this work. A derivation of this expression can be found in Ref.~\cite{Tsog_pre19}. 

\subsubsection{An excess entropy}

For the hard-sphere system in which an internal energy is zero, the excess entropy $S^{e}/Nk_{B}$ can be computed as a following thermodynamic relation \cite{McQuarried10}
\begin{eqnarray}\label{exentropy}
 \frac{S^{e}}{N k_{\mathrm{B}}} = \frac{\beta p}{\rho} -\beta \mu^{e} - 1. 
\end{eqnarray}
In evaluating expression (\ref{exentropy}), we use previously obtained values for the pressure and excess chemical potential.

\section{Results and discussion}

In this work we have done calculations for a reduced densities: $\rho\sigma^{3} = 0.1$ to $0.9$. For high density, such as, when $\rho\sigma^{3} \sim 0.9$, the hard-sphere system behaves like a fluid \cite{Hansen06}. 
We use the Picard iteration method to solve the OZ equation, in which the OZ equation (\ref{oz}) is solved in the Fourier space while the closure equation (\ref{cl}) is solved in a coordinate space. A number of grid points in a length interval of $[0, \, 16\sigma]$ is $2^{15}$. The numerical calculation has been done with an in-house Matlab \cite{Matlab15} code. The numerical tolerance for a root-mean-squared residual of the indirect correlation functions of a successive iterations was set at $10^{-8}$.    

In table I we have compared the values of the equation of states (EOS) and excess chemical potential obtained from both the PY and MS approximations with available data in Ref.~\cite{Zhou06}, 
which was obtained in terms of the Carnahan and Starling (CS) EOS calculation \cite{Carnahan69}. We note that the CS data \cite{Carnahan69} can be considered as `exact' values \cite{Zhou06}.  
\begin{table}[h]
\caption{The $\beta p/\rho$ and $\beta \mu^{e}$ as a function of $\rho \sigma^{3}$. }{%
\begin{tabular}{c c c c c c c c}
\hline \hline 
& \multicolumn{3}{c}{$\beta p/\rho$} & \multicolumn{4}{c}{$\beta \mu^{e}$} \\
\cline{2-8} 
$\rho \sigma^{3}$ & PY & MS & Ref.~\cite{Carnahan69}  & & PY & MS & Ref.~\cite{Carnahan69}  \\
 \hline
 0.1 & $1.24$ & $1.24$ & $1.2397$ & & $0.46$ & $0.46$ & $0.4637$\\
 0.2 & $1.55$ & $1.55$ &  & & $1.03$ & $1.03$  \\
 0.3 & $1.95$ & $1.96$ & $1.9667$ & & $1.72$ & $1.73$ & $1.7468$  \\
 0.4 & $2.48$ & $2.50$ & & & $2.58$ & $2.62$ &  \\
 0.5 & $3.17$ & $3.23$ & $3.2624$ & & $3.67$ & $3.77$ & $3.8068$  \\
 0.6 & $4.09$ & $4.23$ & &  & $5.09$ & $5.27$ \\
 0.7 & $5.32$ & $5.61$ & $5.7102$ & & $7.00$ & $7.32$ & $7.3593$  \\
 0.8 & $7.00$ & $7.58$ & $7.7497$ & & $9.70$ & $10.2$ & $10.1525$ \\
 0.9 & $9.33$ & $10.4$ & $10.7461$ & & $13.7$ & $14.3$ & $14.1052$ \\
\hline \hline
\end{tabular}}{}
\label{tab2}
\end{table}
For higher densities, values for both quantities from PY approximation are lower than those from the MS approximation. The MS values are quite close to those obtained in Ref.~\cite{Carnahan69}.  

Table II presents our findings for an excess entropy $S^{e}/Nk_{\mathrm{B}}$ obtained in both approximations and their comparison with those of Refs.~\cite{Zhou06} and \cite{Carnahan69}. As expected, the largest deviations come from the PY results here, especially for high density, since both the PY pressure and excess chemical potential values are lower (previous table). Values from Ref.~\cite{Zhou06} are obtained with the help of hybrid-bridge function, which enables better values than ours.   
\begin{table}[h]
\caption{An excess entropy $S^{e}/Nk_{\mathrm{B}}$.}{%
\begin{tabular}{c c c c c c c c}
\hline \hline 
& \multicolumn{3}{c}{$S^{e}/Nk_{\mathrm{B}}$} \\
\cline{2-5} 
$\rho \sigma^{3}$ & PY & MS & Ref.~\cite{Zhou06} & Ref.~\cite{Carnahan69}  \\
 \hline
 0.1 &  $-0.22$ & $-0.22$ & $-0.2240$ & $-0.2241$ \\
 0.2 &  $-0.48$ & $-0.48$ &  \\
 0.3 &  $-0.77$ & $-0.77$ & $-0.7790$ & $-0.7801$ \\
 0.4 &  $-1.10$ & $-1.12$ \\
 0.5 &  $-1.50$ & $-1.53$ & $-1.5414$ & $-1.5444$ \\
 0.6 &  $-2.00$ & $-2.04$ \\
 0.7 &  $-2.68$ & $-2.71$ & $-2.6851$ & $-2.6491$ \\
 0.8 &  $-3.70$ & $-3.60$ & $-3.3824$ & $-3.4028$\\
 0.9 & $-5.39$ & $-4.87$ & $-4.3555$ & $-4.3591$\\
\hline \hline
\end{tabular}}{}
\label{tab2}
\end{table}
Based on results shown in tables I and II we note that the MS approximation makes better calculation than the PY does. A reason why this happens might be related to the fact that an correlation function from the MS approximation are better than those from the PY approximation \cite{Banz22, Banz22_num}. Finally, note that our results obtained are independent on number of grid points and a length interval employed. 

\section{Conclusion} 	

In this work we have implemented the Orsntein-Zernike integral equation theory with the Percus-Yevick and Martynov-Sarkisov bridge functions for single-component hard-sphere system. Analytical expressions based on correlation functions which have been obtained as a solution of the integral equation for a density, and an excess chemical potential and entropy have been tested. Our findings for these thermodynamic quantities from both approximations have been compared with available accurate data. It has been shown that Matrynov-Sarkisov approximation does work better than the Percus-Yevick approximation in which the integral equation can have a closed form solution \cite{Laird92}. The MS values are close to the accurate data. Note that it is better to use a well-approximated bridge functions, such as hybrid or pressure consistent bridge approximations \cite{Zhou06} to obtain well-consistent data from the integral equation method.

\section{ACKNOWLEDGEMENT(S)}
This research work has been supported by the Mongolian Foundation for Science and Technology (Project No. ShUTBIKhKhZG-2022/167).


\end{document}